\documentclass[pdflatex,sn-nature,referee,sort&compress]{sn-jnl}

\usepackage{graphicx}%
\usepackage{multirow}%
\usepackage{amsmath,amssymb,amsfonts}%
\usepackage{amsthm}%
\usepackage{mathrsfs}%
\usepackage[title]{appendix}%
\usepackage{xcolor}%
\usepackage{textcomp}%
\usepackage{manyfoot}%
\usepackage{booktabs}%
\usepackage{algorithm}%
\usepackage{algorithmicx}%
\usepackage{algpseudocode}%
\usepackage{listings}%
\usepackage{subcaption}

\theoremstyle{thmstyleone}%
\theoremstyle{thmstyletwo}%

\theoremstyle{thmstylethree}%

\begin{document}

\title[Impact of Stratospheric Aerosol Injection on Renewable Energy Systems]{Impacts of Stratospheric Aerosol Injection on Renewable Energy Systems}

\author*[1]{\fnm{Sebastian} \sur{Kebrich}}\email{s.kebrich@fz-juelich.de}
\equalcont{These authors contributed equally to this work.}

\author*[1]{\fnm{Luisa} \sur{Kamp}}\email{luisa.kamp@rwth-aachen.de}
\equalcont{These authors contributed equally to this work.}

\author[1]{\fnm{Jochen} \sur{Linßen}}\email{j.linssen@fz-juelich.de}

\author[1,3]{\fnm{Heidi} \sur{Heinrichs}}\email{h.heinrichs@fz-juelich.de}

\affil[1]{\orgdiv{Institute of Climate and Energy Research – Jülich Systems Analysis (ICE--2)}, \orgname{Forschungszentrum Jülich GmbH}, \orgaddress{\street{Wilhelm-Johnen Straße}, \city{Jülich}, \postcode{52425}, \state{Nordrhein-Westfalen}, \country{Germany}}}

\affil[2]{\orgdiv{ Chair for Energy Systems Analysis, Department of Mechanical Engineering}, \orgname{University of Siegen}, \orgaddress{\street{Paul-Bonatz Straße 9-11}, \city{Siegen}, \postcode{57076}, \state{Nordrhein-Westfalen}, \country{Germany}}}

\abstract{Climate change is one of the 21st century's major challenges. However, the progress in reducing greenhouse gas emissions is perceived as being too slow. Hence, 
more radical technologies such as stratospheric aerosol injection are entering discussions to limit climate change.
This study presents a methodology for evaluating the effects of injecting 20Mt of SO$_2$ into the atmosphere annually on the global radiative balance, photovoltaic potentials, and renewable energy systems under a targeted temperature reduction of $2^\circ$C. Results show that the average annual reduction of PV potentials ranges from $0.25\%$ to $4\%$ up to $12\%$ in Northern Europe during summer. The modeled renewable energy systems largely absorb these reductions resulting in minor capacity shifts with larger changes confined to a few systems. 
The results show that the inherent flexibility of large--scale renewable energy systems helps mitigating changes in cost, but understanding this flexibility is crucial to avoid errors in design.}


\maketitle

\section{Introduction}\label{sec1}

Global warming, driven by rising concentrations of greenhouse gases in the Earth's atmosphere, is one of the most critical challenges of this century \cite{irena_global_2022}. According to the latest Copernicus Climate Report, the international goal of limiting global warming to between 1.5 and $2^\circ$C above pre-industrial levels was exceeded for the first time in 2024 \cite{copernicus_climate_change_service_c3s_global_2025}. Despite the steady increase in renewable energy technology capacity, with 447 GW of PV deployed in 2024 alone \cite{irena_world_nodate, dunster_global_nodate}, the main source of energy remains conventional energy carriers like oil and gas. This has led to a steady increase in greenhouse gases since most countries plan to be greenhouse gas neutral by mid--century. This development underscores the intensified discourse on alternative approaches to mitigating the effects of global warming.

In this context, solar radiation management technologies are gaining increasing attention as potential tools to reduce the Earth's temperature by enhancing the reflectivity of incident solar radiation \cite{baur_solar_2024}. One such technology is Stratospheric Aerosol Injection, which is expected to have high impact. It is also well understood due to its comparability to volcanic eruptions \cite{huynh_potential_2024}. In this solar radiation management technology, sulfur dioxide ($SO_2$) is injected directly above the troposphere into the stratosphere to scatter incoming solar radiation. This alters the Earth's radiative balance, reducing direct irradiance while increasing diffuse irradiance and resulting in a net cooling effect on the atmosphere \cite{bochuer_7_nodate}.

The cooling potential of $SO_2$ has recently been demonstrated by the impact of reduced emissions from ship tanks. Although these reductions were aimed at improving air quality, they contributed to a warming effect by reducing the reflection of solar radiation \cite{hausfather_analysis_2023}. Despite the demonstrated cooling effects, however, the wider scientific and policy discourse on solar radiation management technologies remains divided. The 2024 expert report on solar radiation management outlines uncertainties and potential risks associated with these technologies, the lack of technology maturity needed, as well as not being suitable as a substitute for greenhouse gas reduction \cite{copernicus_climate_change_service_c3s_global_2025}. Nevertheless, the report emphasizes the need for increased research to fill significant gaps in knowledge, particularly on the direct and indirect effects of solar radiation management.

Previous research on Stratospheric Aerosol Injection has focused on the atmospheric and climatic effects of induced $SO_2$, such as temperature patterns, atmospheric oscillations, and the carbon and hydrological cycle \cite{irvine_overview_2016, huynh_potential_2024, muri_climate_2018, moch_overlooked_2023}. However, recent studies have also investigated secondary effects on social structures \cite{kumler_potential_2025}. Due to the direct relationship between solar irradiance and photovoltaic (PV) performance, a few studies have begun to examine the impact on PV potentials \cite{baur_solar_2024, ojo_impact_2024, smith_impacts_2017}. These studies assume the injection of 5--15Mt $SO_2$ per year to offset 1 to $1.5^\circ$C of global warming and use either the climate model Hadley Centre Global Environment Model \cite{martin2006physical} or the Coupled Model Intercomparison Project Phase 6 \cite{makula2022coupled} for their calculations. These results show a reduction in PV potential between 1\% and 7\% globally \cite{baur_solar_2024, smith_impacts_2017}, but the study on Nigeria reports increased radiation levels across the country except the coastal regions \cite{ojo_impact_2024}. 
However, the possible reduction of PV performance due to Stratospheric Aerosol Injection (SAI) may therefore impact the needed renewable energy technology capacities \cite{baur_solar_2024}. 

Stratospheric aerosol injection was selected based on a better understanding of its impact on different wavelengths, regional distribution, and seasonal dependency compared to other climate engineering technologies \cite{moch_overlooked_2023}. Additionally, its similarity to volcanic eruptions serves as a natural analogy, allowing for better estimation of the impact and support modeling of the technology \cite{huynh_potential_2024}. Furthermore, although the aerosol injection process has to be repeated yearly due to sedimentation, the injection of 20Mt of sulfate precursors is estimated to be around -2.5$\frac{W}{m^2}$ corresponding to a temperature decrease of $2^\circ$C \cite{muri_climate_2018}. The injection process is simulated over a duration of ten years to establish an equilibrium state between injection, distribution and sedimentation. The injection location is along the equator, so that the aerosols can then be carried toward the North and South Poles by natural wind patterns, resulting in a rather uniform distribution worldwide.  The remainder of the study, first examines the impact on the irradiance of stratospheric aerosol injection, then the impact on the resulting PV potentials, and finally the impact on energy systems worldwide, comparing systems that consider and do not consider the impact of SAI on PV potentials. The underlying methodology can be found in the appendix.

\section{Main}\label{sec2}

The analysis covers three aspects. First, the global reduction in the radiative balance is presented. The average annual reduction ranges from 0.8$\frac{W}{m^2}$ in (sub-)tropical regions and 3.1$\frac{W}{m^2}$ beyond $50^\circ\text{N}$ and $50^\circ\text{S}$, reaching up to 10$\frac{W}{m^2}$ during the summer. Second, the impact on PV potentials is examined. The average yearly reduction lies between 2 and 4\% for Northern America, Northern Europe and Northern Asia, but can be as high as 12\% in Northern Europe during summer. Third, the impact on energy systems worldwide is evaluated by optimizing energy systems with and without the impact of SAI on PV potentials. The results show that large--scale energy systems absorb the impact largely, resulting in only minor capacity adjustment. Larger shifts are observed for a few, smaller energy systems.

\begin{figure}[H]
    \centering
    \begin{subfigure}{\textwidth}
        \centering
        \includegraphics[width=\textwidth]{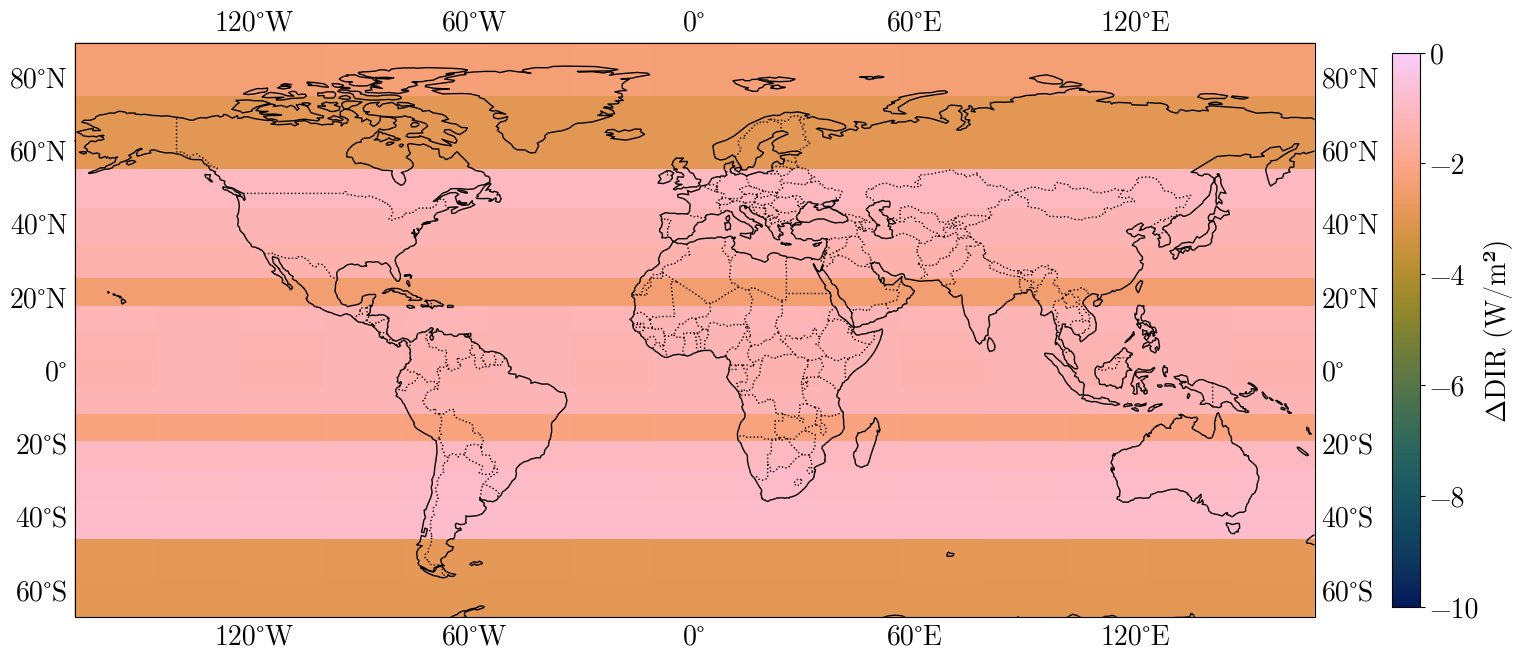}
        \caption{Yearly average reduction in direct irradiance assuming an injection of 20 Mt of sulfate precursors along the equator after ten years of simulated injection.}
        \label{fig:delta_dir_annual}
    \end{subfigure}
    \begin{subfigure}{0.49\textwidth}
        \centering
        \includegraphics[width=\linewidth]{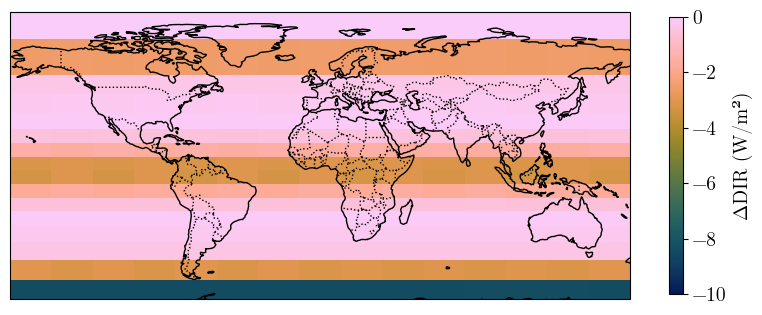}
        \caption{Reduction in direct irradiance (Dec–Feb) after ten years of simulated injection.}
        \label{fig:delta_dir_djf}
    \end{subfigure}
    \hfill
    \begin{subfigure}{0.49\textwidth}
        \centering
        \includegraphics[width=\linewidth]{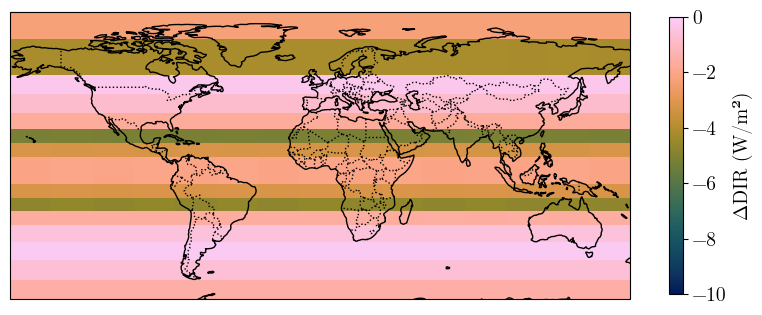}
        \caption{Reduction in direct irradiance (Mar–May) after ten years of simulated injection.}
        \label{fig:delta_dir_mam}
    \end{subfigure}
    \begin{subfigure}{0.49\textwidth}
        \centering
        \includegraphics[width=\linewidth]{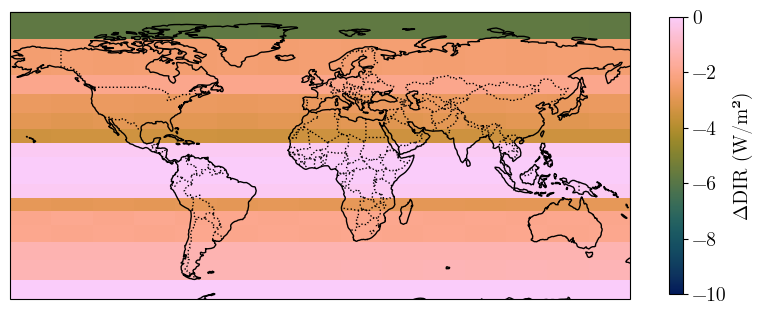}
        \caption{Reduction in direct irradiance (Jun–Aug) after ten years of simulated injection.}
        \label{fig:delta_dir_jja}
    \end{subfigure}
    \hfill
    \begin{subfigure}{0.49\textwidth}
        \centering
        \includegraphics[width=\linewidth]{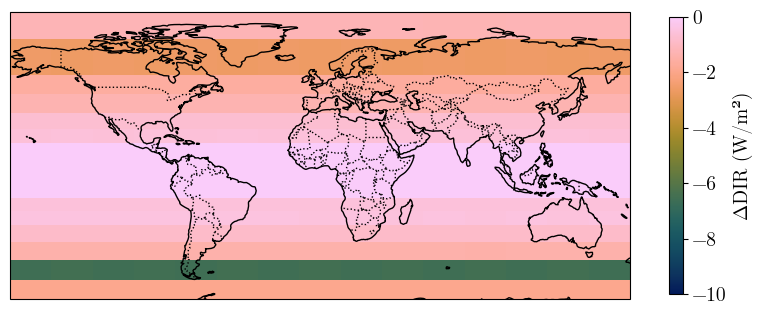}
        \caption{Reduction in direct irradiance (Sep–Nov) after ten years of simulated injection.}
        \label{fig:delta_dir_son}
    \end{subfigure}

    \caption{
        The reduction in direct irradiance is shown annually as well as seasonally after simulated injection of 20Mt of SO$_2$ in December annually for ten consecutive years to reach an equilibrium state between injection, distribution and sedimentation. Panel (a) shows the annual average reduction in direct irradiance (\(\Delta\)DIR), ranging from 0.8$\frac{W}{m^2}$ in the tropics to 3.1$\frac{W}{m^2}$ starting around 50°S and 50°N. Around 20°S and 20°N an increased reduction of 2$\frac{W}{m^2}$ arises from a weak tropical to subtropical exchange. Panels (b–e) show the seasonal mean changes in the same scenario. Reduction is high in the northern and southern hemisphere during the respective summer reaching up to 10$\frac{W}{m^2}$. In the tropics the reduction is strongest directly in the month after injection in December with around 4$\frac{W}{m^2}$ as shown in Panel (b). In the month from March to May shown in Panel (c) the reduction weakens and the belts around 20°S and 20°N form. In the remaining month from June to November the reduction in the tropics is negligible. The reduction starting around 20°S and 20°N is between 0.3$\frac{W}{m^2}$ and 7.8$\frac{W}{m^2}$ strongest during the summer in the month from June to November.
    }
    \label{fig:Delta_dir_maps}
\end{figure}

\textbf{Radiative Balance Changes are strongest during summer and beyond $50^\circ$ North and South}

Not only does the regional reduction of the radiative balance vary on average, it also varies greatly across seasons. Figure \ref{fig:Delta_dir_maps} shows the mean annual reduction as well as the average seasonal reduction in direct irradiance ($\frac{W}{m^2}$) worldwide. The most affected areas include Northern Europe, Russia and parts of Canada, where annual average losses reach up to 3.1$\frac{W}{m^2}$ compared to the case without SAI. This effect is similar to the global radiative forcing that the modeled SAI scenario aims to offset (3.93$\frac{W}{m^2}$). Similar intensities are observed in regions below 50°S, but, fewer countries are affected due to less land mass. The global spatial pattern is shaped by the dominant atmospheric circulations from the equator toward the poles. Although the aerosol injection point is near-equatorial to use the natural transport patterns from the equator the poles, a relatively weak exchange between tropical and extratropical regions leads to an accumulation in mid-latitudes. This results in characteristic belts of reduced irradiance around 20°N and 20°S, with typical average reductions of about 2$\frac{W}{m^2}$ in direct irradiance, as shown in Figure \ref{fig:Delta_dir_maps}. SAI's effect on direct irradiance shows strong seasonal variations. Stronger reductions occur in regions below and above $50^\circ$S and $50^\circ$N during their respective spring and summer seasons as shown in Figure \ref{fig:Delta_dir_maps}, b-e. Tropical regions experience only minor reductions throughout the year, except near 20°N/S, where localized peaks occur in spring (March–May) due to aerosol transport patterns where the average irradiance reduction reaches up to –6$\frac{W}{m^2}$. Nonetheless, besides the accumulation between the tropics and the extratropics, the impact on the radiation in the tropics is around -0.8$\frac{W}{m^2}$. \\

\textbf{Reduction of PV Potentials strongest during summer in the North}\\

SAI-induced changes in the radiative balance result in a consistent global decrease in PV capacity factors that closely mirror the spatial patterns of direct irradiance losses. The strongest negative effects are observed in Norther America, Northern Europe, and Northern Asia, where aerosol transport toward the poles and increasing atmospheric variability lead to significant decreases. During summer, Northern Europe experiences a reduction of up to 12\%, while reductions are between 2\% and 4\% during the remaining seasons. In Canada and Russia, reductions reach up to 6\% during the summer and range from 2\% to 4\% throughout the rest of the year. The rest of the world experiences an average PV potential reduction of less than 1\% on average, reaching around 2\% in the area between the tropics and extratropics near 20°N/S during spring due to the accumulation of aerosols caused by the weak exchange between the two regions. 
    
\begin{figure}[H]
    \centering
    \begin{subfigure}{\textwidth}
        \centering
        \includegraphics[width=\textwidth]{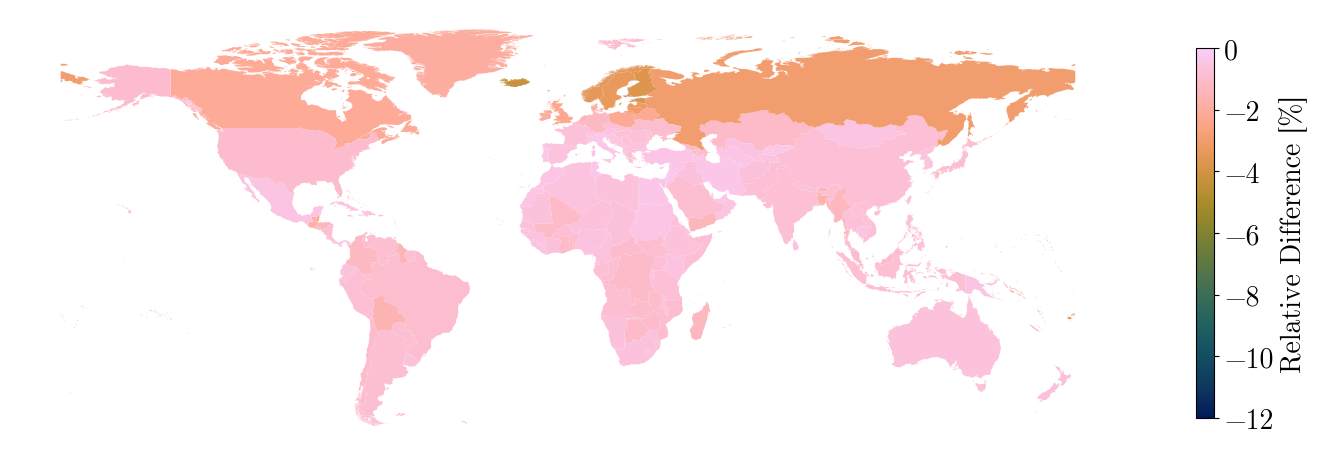}
        \caption{Yearly average reduction in PV capacity factors assuming an injection of 20 Mt of sulfate precursors along the equator after ten years of simulated injection.}
        \label{fig:meancapfac_annual}
    \end{subfigure}
    \begin{subfigure}{0.49\textwidth}
        \centering
        \includegraphics[width=\linewidth]{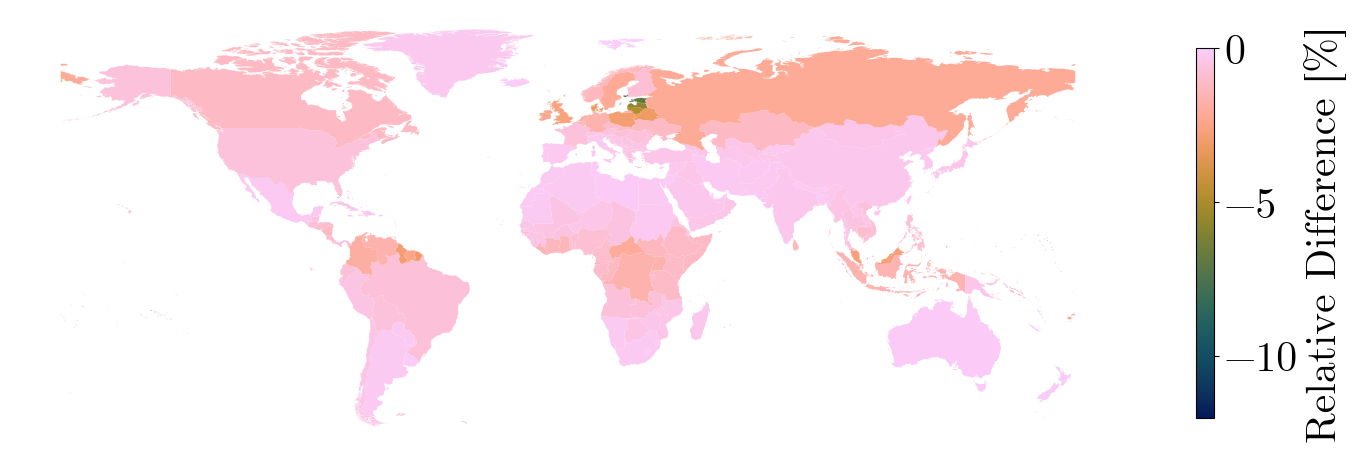}
        \caption{Reduction in PV capacity factors (Dec–Feb) after ten years of simulated injection.}
        \label{fig:meancapfac_djf}
    \end{subfigure}
    \hfill
    \begin{subfigure}{0.49\textwidth}
        \centering
        \includegraphics[width=\linewidth]{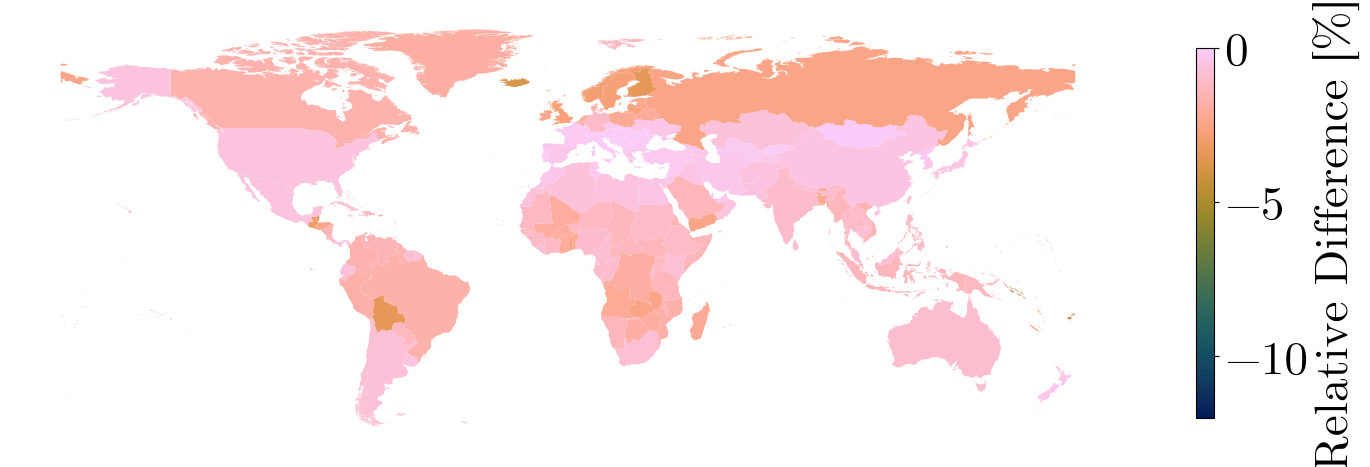}
        \caption{Reduction in PV capacity factors (Mar–May) after ten years of simulated injection.}
        \label{fig:meancapfac_mam}
    \end{subfigure}
    \begin{subfigure}{0.49\textwidth}
        \centering
        \includegraphics[width=\linewidth]{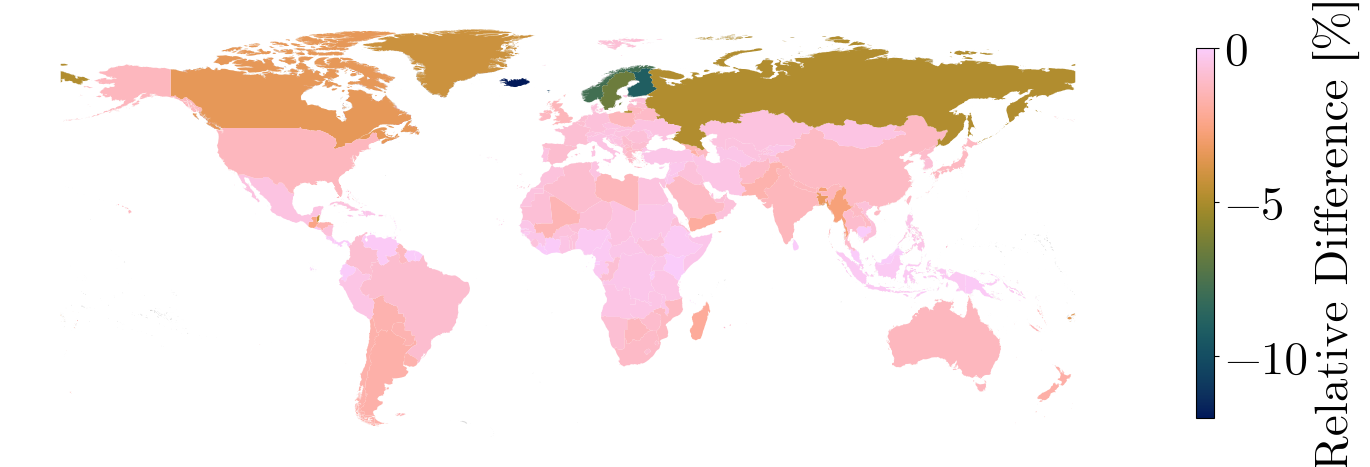}
        \caption{Reduction in PV capacity factors (Jun–Aug) after ten years of simulated injection.}
        \label{fig:meancapfac_jja}
    \end{subfigure}
    \hfill
    \begin{subfigure}{0.49\textwidth}
        \centering
        \includegraphics[width=\linewidth]{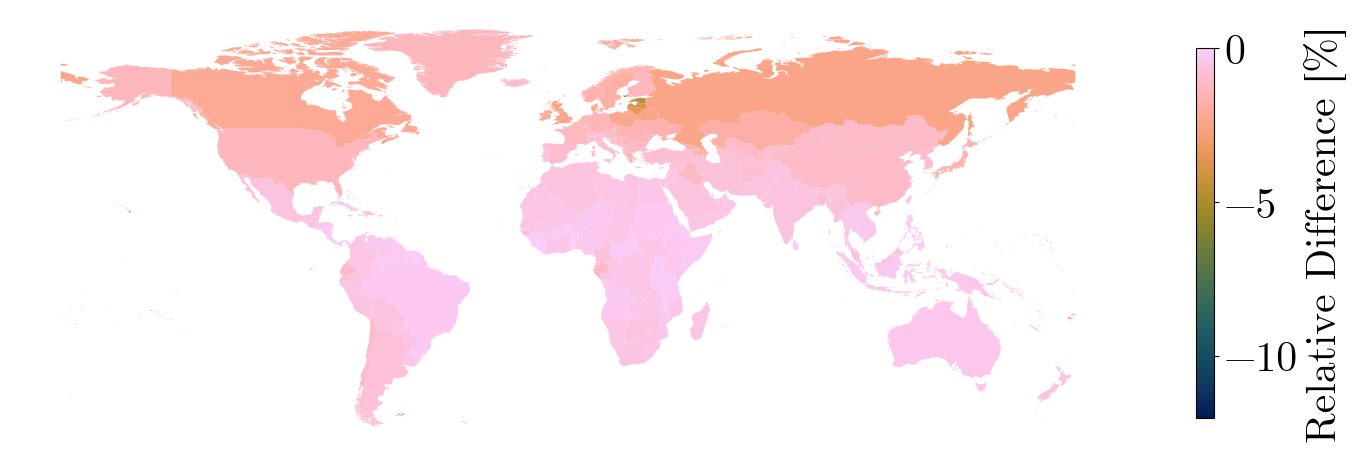}
        \caption{Reduction in PV capacity factors (Sep–Nov) after ten years of simulated injection.}
        \label{fig:meancapfac_son}
    \end{subfigure}

    \caption{
        The relative difference in PV capacity factors is shown annually as well as seasonally equivalently to the reduction in direct irradiance in Figure \ref{fig:Delta_dir_maps}. Panel (a) shows the relative annual average relative difference in PV capacity factors, ranging from 0.25\% in the tropics to 4\% Northern Europe. Panels (b–e) show the seasonal relative difference in the same scenario. Reduction is high in the northern hemisphere during the summer reaching up to 12\% in Northern Europe. In the tropics the relative difference reaches up to 2\% in the month after injection shown in Panel (b). In the month from March to May shown in Panel (c) the relative difference in the tropics weakens and instead spreads out to the subtropics. In the month from June to November the relative difference in the tropics reduces further and is negligible in the month from September to November. The relative difference in Northern America, Northern Europe and Northern Asia is similar throughout the year besides the summer, where it reaches up to 12\%.
    }
    \label{fig:map_capfacs}
\end{figure}

\textbf{Flexibility in Design of large--scale Renewable Energy System Mostly Compensate Impacts of SAI}

The total annual cost of energy systems worldwide remains effectively unchanged (-0.25\%-- +0.5\%), but shifts in capacity per technology can be identified. In most regions, the relative differences in capacity are below 10\%, and especially large energy systems in terms of area and demand both — such as those covering North America, Europe, Northern Asia, China, and India — show only marginal adjustments. Larger shifts (10-20\%) occur in four out of the 28 unions, while for one of the modeled unions,  the overall design of its energy system changes more significantly.\\

\noindent
\emph{Baseline renewable energy system configuration}

Changes in photovoltaic (PV) potentials resulting from stratospheric aerosol injection were assessed for fully renewable energy systems worldwide. Energy systems were optimized for 28 unions under two scenarios: one with standard PV potentials worldwide and one PV potentials that include SAI--induced changes. All other parameters were held constant to isolate the impact of modified PV potentials on the overall system design. A brief model description can be found in the appendix, and a full description, including all union definitions, techno-economic parameters, and potentials, has been published by Franzmann et al. \cite{franzmann2025impact}.\\

\begin{sidewaysfigure}
    \includegraphics[width=\textwidth]{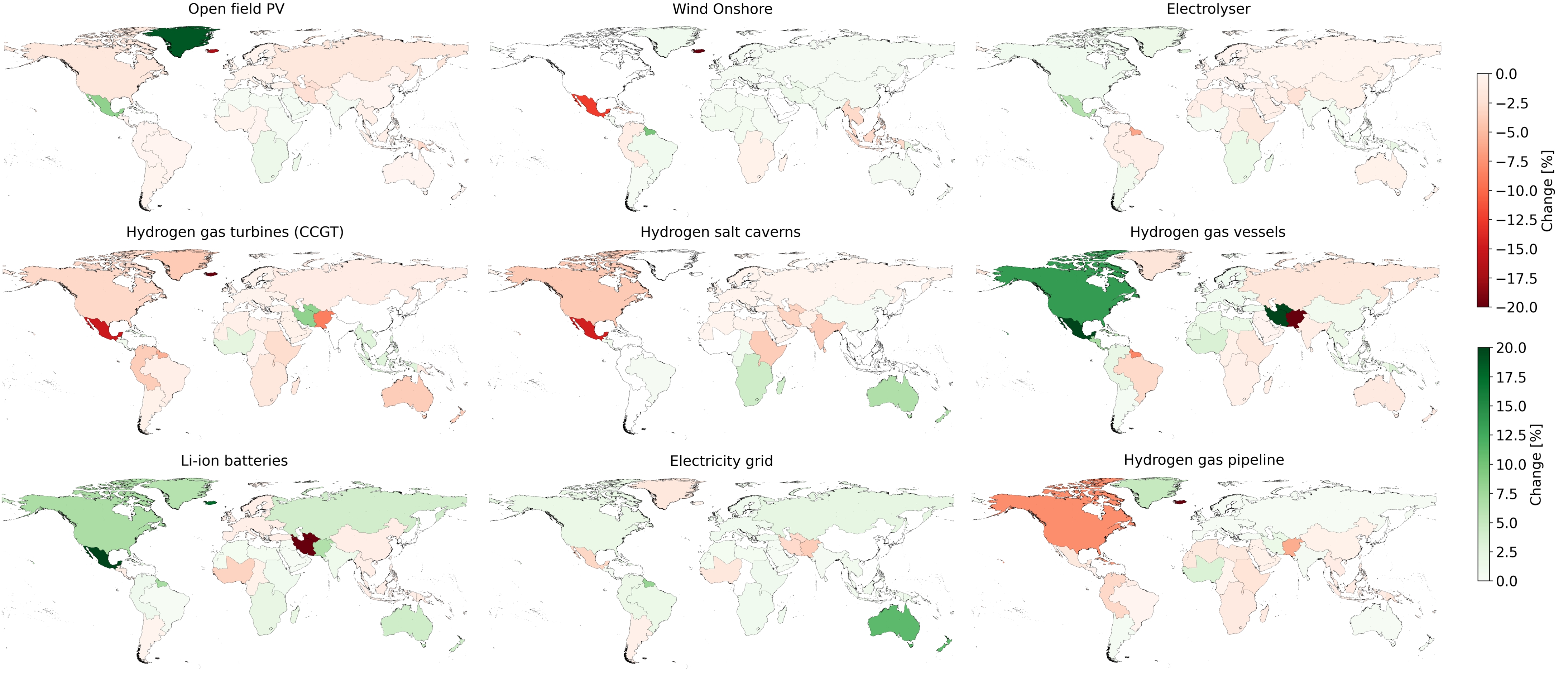}
    \caption{The relative difference in the installed capacities of energy systems worldwide, split into 28 unions cut off at 20\% is shown. Geothermal energy is not depicted because the changes are negligible in all but one modeled union (Mexico, ). Changes are small for most unions (<10\%), especially for larger energy systems covering North America, Europa, and Northern Asia as well as China and India. These systems show a slight shift from PV and long--term storage using hydrogen towards wind energy. There is an additional increase in transmission and short-term flexibility based on a slight increase in electricity transmission combined with an increase in either hydrogen vessels or Li--ion batteries. Larger shifts (10--20\%) are observed in only five of the 28 unions, one of which (Mexico) shows significant changes in design. Greenland, Iceland, Afghanistan, and Pakistan, as well as Turkmenistan, Iran, and Armenia show larger shifts (10-20\%) in the relative shares of PV, wind, hydrogen gas turbines, hydrogen vessels, and Li-ion, respectively, but these technologies' overall share of operation is small, making up less than 10\% of the energy system. Mexico's open field PV shows an increased by 9\%, similar to geothermal, which increases by 12\%, while wind decreases by 12\%. Furthermore, short--term storage flexibility increases for hydrogen vessels (+36\%) and Li--ion batteries (+165\%), while long--term storage in form of hydrogen gas turbines (-15\%) and hydrogen salt caverns (-15\%) decreases.}
    \label{fig:reldiffESM}
\end{sidewaysfigure}

Figure \ref{fig:reldiffESM} shows these changes in capacity. Overall, several shifts are recognizable worldwide, with a few exceptions. As expected, there is a shift away from PV and toward wind, especially in North America, Europe, and Northern Asia, where the PV potential is reduced the most. Short-term storage increases, either via hydrogen gas vessels or lithium-ion batteries, while long-term storage relying on hydrogen decreases. Furthermore, electricity transmission increases worldwide.

Similar to the changes in PV potential shown in Figure \ref{fig:map_capfacs}, where North America, Northern Europe and Northern Asia have the highest changes of up to 4\%, the reduction in PV capacity is up to 2\%. Countries in northern Europe, for example, rely more on energy from wind and geothermal, as well as on energy imports from PV in southern Europe, and are therefore less impacted. Greenland and Iceland, both island systems, demonstrate contrasting trends, with Greenland increasing its capacities for both PV and wind, while Iceland is reducing both. Iceland can rely on its vast geothermal potential; the country already relies on this resource, which provides about 95\% of its energy, even without SAI. Greenland's energy system does not have this option and must therefore increase PV and wind capacities to compensate for reduced PV yield.

Two more modeled unions that show shifts larger than 10\% are the union comprising Afghanistan and Pakistan and the union comprising Turkmenistan, Iran and Armenia. Both show similar shifts away from PV and towards wind energy. However, the former one increases its reliance on hydrogen gas turbines and vessel, while the latter one increased its reliance on Li--ion batteries.

An exceptionally large change occurs in the design of the Mexican energy system. Although the average capacity factor decreases by only 0.5\% similar to the increase in total annual cost by 0.5\%; the significant change in design indicates a high sensitivity to a reduction in its PV potential. Open field PV and geothermal capacities increase by 9\% and 12\%, respectively, while wind decreases by 12\%. The change in capacities is even more pronounced for storage, the hydrogen sector, and transmission. Short--term storage increases for hydrogen vessels (+36\%) and Li--ion batteries (+165\%) while long--term storage in the form of hydrogen gas turbines (-15\%) and hydrogen salt caverns (-15\%) decreases, and electrolyzers (+6\%) which for all other unions only shows minor changes.

In summary, the flexibility of energy systems, based on their combination of different technologies and large interconnected regions, mostly prevents cost changes. Some trends can be identified, such as the tendency to reduce PV, which aligns with the reduction of PV potentials. Only the Mexican energy system reacts with large-scale design changes, even though its PV potential reduction is minor.

\section{Discussion}\label{sec12}

Although it has been demonstrated that SAI reduces surface solar irradiance and PV, as well as CSP potentials, its implications for energy systems have been largely unexplored. This study investigates the impact of SAI on PV potentials and on fully renewable energy systems worldwide. To the best of our knowledge, these investigations into energy systems are novel and extend existing literature by quantifying the necessary changes in their design.

The reduction in irradiation and corresponding PV potential is most notable in high-latitude regions. The most severe reductions in direct irradiation, up to -10$\frac{W}{m^2}$, occur during the summer in both the Northern and Southern Hemispheres, which reduces the impact on energy systems. The average reduction of the global PV potential ranges from 0.25\% in tropical regions to 4\% observed in areas such as North America, northern Europe, and northern Asia. Similarly, the reduction in PV potential is strongest beyond 50$^\circ$N and 50$^\circ$S during the summer when the high availability can absorb the reductions well. The results of the two global studies in literature show a reduction in from 1\% to 3\% for the PV potential for an intended 1°C reduction \cite{smith_impacts_2017} and from 1.4\% to 6.9\% for the PV potential for an intended 1.5-2°C reduction \cite{baur_solar_2024}. These results overlap in large parts with the results of 0.25\% to 4\% for this study with an intended temperature reduction of 2°C. The novel results on optimized energy systems based on renewable energy technologies worldwide demonstrate how flexibility in design, i.e. installed capacities, allows to absorb these impacts resulting in negligible changes in total annual cost (-0.25\%-- +0.5\%).

Although the expected impact on PV's dominant role in energy systems is limited, consistent regional disparities drive subtle yet measurable shifts in the energy mix of these systems. The inherent flexibility of large-scale energy systems — capable of balancing generation from diverse sources — largely mitigates effects on total annual system costs. At the global scale, reduced irradiance and PV potentials slightly favor wind power expansion, indicating a marginal but systematic shift away from PV-dominated configurations toward more wind-based systems.  

However, national systems with lower flexibility are more sensitive to these perturbations. Countries must therefore recognize the structural properties of their energy systems and their capacity to adapt to irradiance variability when designing long-term energy strategies. Mexico exemplifies this sensitivity: despite only minor reductions in PV potential, its optimal system configuration changes markedly, underscoring how small shifts in solar resources can propagate through system design choices. While in this study, the modeled unions are mostly large-scale energy systems, real-world transitions may lead to more distributed and regionally autonomous configurations, likely enhancing the observed shifts.

The scope of this work is limited to the effects of SAI on PV potentials and its impact on renewable energy systems. SAI likely impacts other energy sources, such as wind, biomass and hydro power, which in turn impacts energy systems based on these technologies. Beyond its impact on energy systems, the consequences of increasing CO$_2$ levels themselves in the atmosphere and the oceans are not offset, especially if SAI discourages decision-makers from reducing CO$_2$ output. The consequences of stopping SAI (termination shock), resulting in a short--term temperature increase similar to the reduction caused by SAI, are not part of this work, but need to be carefully considered to avoid unwanted and potentially severe consequences.

Future research on the impacts of SAI on energy systems should focus on energy systems that are either regionally constrained or have limited demand as both limit their inherent flexibility. Further, it is also needed to quantify how SAI affects other renewable energy potentials like wind and hydropower, and how these combined changes impact the energy system configuration.

\section*{Acknowledgements}

This work was partly supported by the Helmholtz Association as part of the Platform for the Design of a Robust Energy System and Raw Material Supply (RESUR) and the program, “Energy System Design.” \\

This work was partly funded by the European Union (ERC, MATERIALIZE, 101076649). Views and opinions expressed are however those of the authors only and do not necessarily reflect those of European Union or the European Research Council Executive Agency. Neither the European Union nor the granting authority can be held responsible for them.

\section*{Declaration of generative AI and AI-assisted technologies in the writing
process}
During the preparation of this work the authors used Perplexity AI, ChatGPT (GPT-4) and DeepL Write in order to improve clarity and language. After using this tool/service, the authors reviewed and edited the content as needed and take full responsibility for the content of the published article.

\section*{Code availability}

The python code is to calculate the changes in radiative balance and the PV potentials is available on github at \hyperlink{https://github.com/FZJ-IEK3-VSA/RESKit/tree/Reskit---Stratospheric-Aerosol-Injection}{Reskit - Stratospheric Aerosol Injection}. A full description of the energy system model used can be found in Franzmann et al. \cite{franzmann2025impact}.

\section*{Declarations}

\noindent
\textbf{Sebastian Kebrich:} Conceptualisation, Methodology, Validation, Formal analysis, Investigation, Writing -- Original Draft, Writing -- Review \& Editing, Visualisation\\
\textbf{Luisa Kamp:} Methodology, Software, Validation, Formal analysis, Investigation, Writing -- Original Draft, Visualisation , Data Curation\\
\textbf{Jochen Linßen:} Resources, Writing -- Review \& Editing\\
\textbf{Heidi Heinrichs:} Conceptualisation, Writing -- Review \& Editing, \\ Supervision, Funding acquisition\\

\section*{Appendix}

\section{Online Methods}

\textbf{Aerosol Distribution}\\

\begin{figure}[t]
    \centering
    \includegraphics[width=\textwidth]{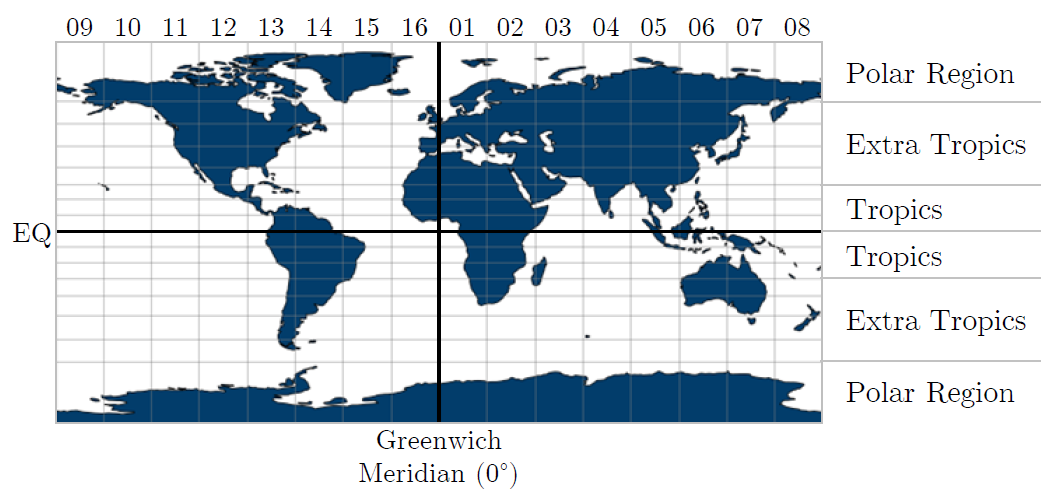}
    \caption{Division of the world into a grid structure. The 16 latitude belts are divided into 16 latitude regions, giving a total of 256 sections. The numbering of the latitudes begins at the Greenwich meridian, the meridian of longitude zero.} 
    \label{fig:worldgrid}
\end{figure}

SAI is modeled within the stratosphere using an atmospheric model that simulates stratospheric motions and aerosol dispersion. The methodology from Gao et al. \cite{gao_volcanic_2008} is applied, focusing on the latitudinal aerosol distribution driven by the Brewer-Dobson circulation from the tropics to the poles. The stratosphere is divided into 16 latitudinal belts: three tropical, four extratropical, and one polar belt in each hemisphere. Each belt is subdivided into 16 sections, yielding a 16x16 spatial resolution in Figure \ref{fig:worldgrid}.

\noindent

\begin{figure}[t]
    \centering
    \includegraphics[width=0.6\columnwidth]{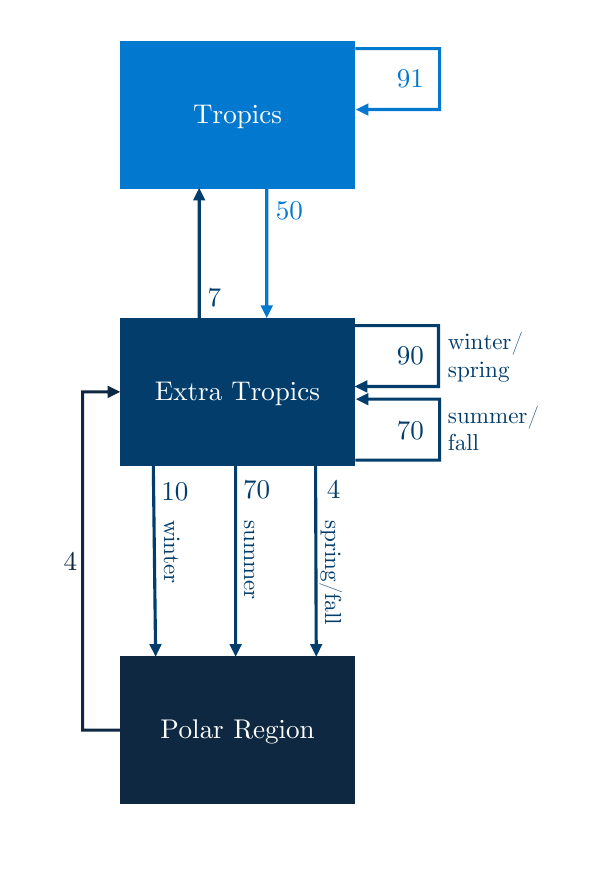}
    \caption{The exchange process used in this work is depicted as described by Gao et al. \cite{gao_volcanic_2008}. Loop arrows indicate the exchange between two belts in the same region. Arrows between blocks indicate the exchange between neighboring belts that are not in the same region.} 
    \label{fig:pub_ex_coeff}
\end{figure}

Within each belt, the aerosol distribution is assumed to be latitudinal homogeneous, with the aerosol optical depth varying with latitude but not with longitude. Aerosol exchange between these belts is determined by seasonal transport coefficients, as shown in Figure \ref{fig:pub_ex_coeff}, describing the aerosol exchange between the respective regions. Most of the movement occurs from the equator to the poles, with minimal transport in the opposite direction.

\noindent

Seasonal exchange matrices (three-month periods) are used to incorporate temporal variability. In addition, the aerosol lifetime is modeled with an e-folding decay rate based on observations \cite{gao_volcanic_2008}. The amount of SO$_2$ injected is assumed to be a linear aerosol build-up over 4 months at the injection point, imitating the effect of a strong volcanic eruption.  After the build-up, the exchange matrix is applied, with region-specific e-folding lifetimes: 36 months for the tropics, 12 months for extratropical regions, and 3 months for polar regions, where removal is fastest due to the Brewer-Dobson circulation \cite{gao_volcanic_2008}.\\


SAI is based on the chemical reaction of SO$_2$ with OH and H$_2$O to form sulfate aerosols. The amount of aerosols, and thus the aerosol optical depth, resulting from a given amount of injected SO$_2$ is calculated using the approach outlined by Metzner et al. \cite{metzner_radiative_2014}. This method provides a reliable parameterization of the global average aerosol optical depth following volcanic eruptions.

The aerosol model assumes a homogeneous distribution of aerosols within each latitudinal belt in megatons per square meter. The total amount of sulfate aerosol for each belt is then determined, resulting in a global average aerosol optical depth. The calculated aerosol optical depth is provided as a time and location-dependent distribution, with monthly values interpolated to produce hourly aerosol distributions.\\

\textbf{Radiative Flux}\\

The effects of the aerosol optical depth distribution on the radiative flux reaching the Earth's surface are assumed to increase the diffuse and decrease the direct radiative component. Consequently, the Delta-Eddington approximation is applied to calculate the radiative fluxes for the direct and diffuse components separately. This approximation aims to model the radiative transfer in an absorbing and scattering atmosphere. It is based on the asymmetry factor and the forward scattering fraction, which represent the optical behavior of aerosols \cite{joseph_delta-eddington_1976}. Due to its simplicity and accuracy, it is widely used in climate models \cite{neale_description_nodate, briegleb_deltaeddington_1992}.

The approximation divides the atmosphere into vertical layers and computes the upward and downward radiative fluxes across each layer interface, where reflectivity and transmissivity for direct and diffuse radiation are determined. For the solution of a single layer, the main input parameters are the scaled properties for the single scattering albedo \(\omega^*\), the aerosol optical depth \(\tau^*\), and the asymmetry factor \(g^*\) \cite{neale_description_nodate}. As a result, the reflectivity and transmissivity to direct radiation (\(R_\text{dir}\), \(T_\text{dir}\)) and to diffuse radiations (\(R_\text{diff}\), \(T_\text{diff}\)) can be calculated for \(\omega < 1\) where \(\mu_0\) is the cosine zenith angle.\\

In this study, the approximation is implemented as a single-layer model for the stratosphere, where the solar constant (S$_0$ = 1361 W/m²) is incident at the top of the atmosphere. 

\begin{figure*}[t!]
    \centering
    \includegraphics[width=0.99\textwidth]{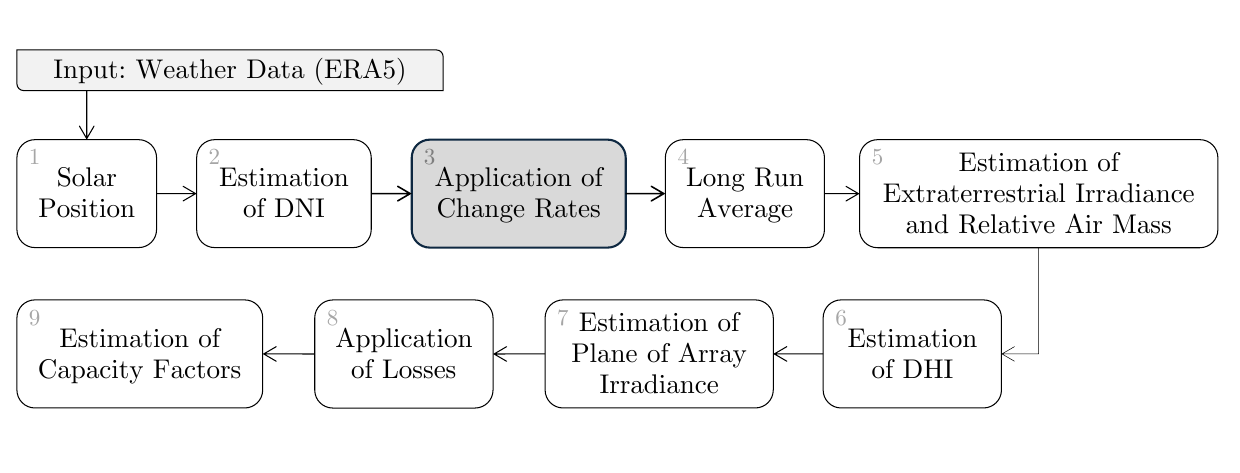}
    \caption[Adapted RESKit Workflow with radiative Change Rates]{Adapted workflow to determine the PV potential in RESKit by applying the change rate in step three.}
    \label{fig: reskit_mod}
\end{figure*}

As a result, the stratospheric layer's reflectivity and transmissivity are influenced only by the sulfate aerosols and do not include any direct interaction with the troposphere.

Following the assumptions made for volcanic aerosols in \cite{neale_description_nodate}, volcanic aerosols are composed of 75\% sulphuric acid and 25\% water, the size distribution is log-normal with an effective radius of 0.426$\mu m$ and a standard deviation of 1.25. The definition of the parameters can be found in the supplementary data of Toohey et al. \cite{toohey_easy_2016}. As the calculation is highly dependent on the solar zenith angle, the python package PV-Lib is used. The package is designed to simulate the performance of PV modules. The aim is to obtain the corresponding zenith angles for any combination of location and time. For each of the 256 sections, the mean value of latitude and longitude is used to calculate the solar zenith angle for each hour \cite{anderson_pvlib_2023}.

According to the radiative balance, the components Direct Horizontal Irradiance (DIR) and Global Horizontal Irradiance (GHI) result in:

\begin{align}
DIR &= DNI \cdot \mu_0\\
GHI &= DIR + DHI
\end{align}

In the Delta-Eddington approximation, calculated directly transmitted flux \(F_{\text{dir}}\) is equal to the definition of Direct Normal Irradiance (DNI). The diffusely transmitted radiation \(F_{\text{diff}}\) reaching the lower boundary corresponds to the definition of Diffuse Horizontal Irradiance (DHI). They are calculated by:
    
\begin{align}
DNI &= F_\text{0} \cdot e^{-\tau^*/\mu_0}\\
DHI &= F_\text{0} \cdot (T_\text{dir} - e^{-\tau^*/\mu_0}) 
\end{align}

To compute the changes in radiative flux due to aerosols, two states of the stratosphere are compared: with and without sulphate aerosols. For the no-aerosol case, DNI equals the solar constant and DHI is zero. For diffuse radiation, a cloud correction factor \(f_\text{corr}\) is applied, incorporating the diffuse fraction coefficient \(k\), which gives information on the clearness of a specific day. Consequently, the scattering effects of clouds in the troposphere are included. The changes in radiative flux are calculated as:

 \begin{align}
    \Delta DNI &= DNI_\text{aerosols} - F_\text{0}\\
    \Delta DHI &= f_\text{corr}(k) \cdot DHI_\text{aerosols}\\
    \Delta DIR &= DIR_\text{aerosols} \cdot \mu_0 - F_\text{0} \cdot \mu_0\\
    \Delta GHI &= F_\text{0} \cdot \mu_0 - DNI_\text{aerosols} \cdot \mu_0 - DHI_\text{aerosols}
\end{align}\\

The paper analyzes the scenario 10 years after the first injection. This allows for a build-up of baseline concentrations in the stratosphere.\\

\textbf{PV Potential Modification in RESKit}\\

To incorporate the influence of SAI on PV potentials throughout changes of the direct and diffuse irradiances, the workflow of the Renewable Energy Simulation Toolkit RESKit is modified. RESKit is a powerful tool for large-scale simulation of renewable energy systems, specifically designed to provide input to energy system models \cite{ryberg_future_2019}.\\

After the estimation of the DNI within the RESKit workflow, as shown in Figure \ref{fig: reskit_mod}, the change rates based on latitude, longitude, and time are applied to the different radiation types. \\
    
The DNI is modified directly by subtracting the change rate \(\Delta DNI\). For the GHI, the change rate is not directly applicable due to its dependence on the DHI, which is not yet determined within the workflow. Consequently, the GHI is separated into its direct and diffuse components, with the \(\Delta DNI\) applied to the DHI and calculated as temporary Diffuse Horizontal Irradiance. In addition, the correction factor \(f_\text{corr}\) is applied to the DHI change rate.

 \begin{align}
    DNI_\text{mod} &= DNI - \Delta DNI \\
    DHI_\text{temp} &= GHI - DIR\\
    GHI_\text{mod} &= (DIR - \Delta DIR) + (DHI_\text{temp} - \Delta DHI \cdot f_\text{corr})
\end{align}\\

Finally, the modified GHI and DNI are used to calculate the adjusted DHI:  
\begin{equation}
    \label{eq:LRA_DHI}
    DHI = GHI_\text{mod} \cdot LRA_\text{GHI} - DNI_\text{mod} \cdot LRA_\text{DNI} \cdot \cos(\alpha_{\text{zenith}}) 
\end{equation}

\begin{figure*}[t!]
    \centering
    \includegraphics[width=0.99\textwidth]{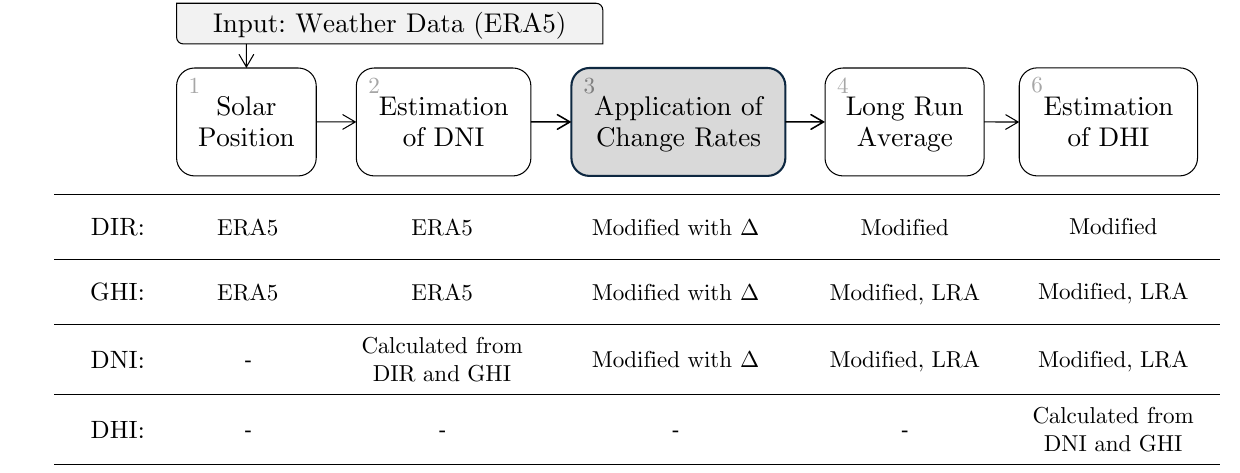}
    \caption[Summary Modifications within the RESKit Workflow]{Summary of the steps where radiative fluxes are calculated or modified. The table describes the status of DIR, GHI, DNI and DHI in each step. (\textit{ERA5} = input values, \textit{Modified by \(\Delta\)} = application of change rates, \textit{LRA} = application of long-term average)}
        \label{fig: reskit_sum}
\end{figure*}

A summary of the workflow modifications is presented in Figure \ref{fig: reskit_sum}, detailing the calculation and adjustment of fluxes at each step.\\

\textbf{Energy System Modeling in ETHOS.FINE}\\

The world is divided into 28 independent energy systems, each optimized individually without cross-border exchange. Each system is modeled using the ETHOS.FINE framework \cite{klutz2025ethos, franzmann2025impact}, incorporating region-specific demand, renewable resource availability, and techno-economic constraints. Energy supply is based on PV, onshore wind, and enhanced geothermal systems \cite{franzmann2025global, ryberg2018evaluating, franzmann2025energy, ishmam2024mapping}. Demand includes both electricity and hydrogen, with energy stored in lithium-ion batteries or hydrogen storage (vessels and salt caverns). Conversion technologies include electrolyzers and hydrogen-fired combined cycle gas turbines, with transmission infrastructure limited to electricity grids and hydrogen pipelines. 

\end{document}